\documentclass[12pt,preprint]{aastex}

\def\a{\mbox{AE\,Aqr}}
\def\e{et~al.\ }
\def\es{{\rm erg\,s^{-1}}}

\newcommand{\be}{\begin{equation}}
\newcommand{\ee}{\end{equation}}
\newcommand{\bdm}{\begin{displaymath}}
\newcommand{\edm}{\end{displaymath}}

\begin{document}

\slugcomment{ApJ Letters, 576, L57 (2002)}

\title{Can the rapid braking of the white dwarf in AE~Aquarii
be explained in terms of the gravitational wave emitter
mechanism\,?}

  \author{Nazar R. Ikhsanov\altaffilmark{1,2} and
Nina G. Beskrovnaya\altaffilmark{2,3}}

  \altaffiltext{1}{Max-Planck-Institut f\"ur Radioastronomie, Auf
dem H\"ugel 69, \\ D-53121 Bonn, Germany
(ikhsanov@mpifr-bonn.mpg.de)}
   \altaffiltext{2}{Central Astronomical Observatory, Russian Academy
   of Sciences, 65-1 Pulkovo, 196140 St.\,Petersburg, Russia}
   \altaffiltext{3}{Instituto Isaac Newton of Chile, St.Petersburg
Branch}

\begin{abstract}
The spin-down power of the white dwarf in the close binary
AE~Aquarii significantly exceeds the bolometric luminosity of the
system. The interpretation of this phenomenon in terms of the
gravitational-wave emitter mechanism has been recently suggested
by Choi \& Yi. The basic assumption of their interpretation is
that the spatially limited blobs or mounds of the mass $ \delta m
\sim 10^{-3}\,M_{\sun}$, are present at the magnetic poles of the
white dwarf. We show that the mounds of this mass can be confined
by the magnetic field of the white dwarf only if the dipole
magnetic moment of the star exceeds $4\times 10^{37}\,{\rm
G\,cm^3}$. Under these conditions, however, the magnetodipole
losses of the white dwarf would exceed the evaluated spin-down
power 6 orders of magnitude. On this basis we discard a
possibility that the observed rapid braking of the white dwarf in
\a\ can be explained in terms of the mechanism proposed by Choi \&
Yi.
\end{abstract}


\section{Introduction}

Rapid braking of the white dwarf in AE~Aquarii is one of the most
puzzling properties of this low-mass close binary system [for the
system parameters and corresponding references see Table~1 in
Ikhsanov \cite{i00}]. As shown by de Jager \e \cite{jmor94}, the
white dwarf is steady spinning down at a rate $\dot{P}=5.64\times
10^{-14}\,{\rm s\,s^{-1}}$, which implies the spin-down power of
   \be\label{lsd}
L_{\rm sd} \cong 6 \times 10^{33}\  I_{50}\ \dot{P}_0\
P_{33}^{-3}\ \es.
  \ee
Here the parameters $I_{50}$, $P_{33}$, and $\dot{P}_0$ denote the
moment of inertia, the spin period, and the spin-down rate of the
white dwarf expressed in units of $10^{50}\,{\rm g~cm^{2}}$,
33\,s, and $5.64 \times 10^{-14}\,{\rm s\,s^{-1}}$, respectively.

$L_{\rm sd}$ exceeds the luminosity of the system in the UV and
X-rays by a factor of 120, and it even exceeds the bolometric
luminosity by a factor of more than 5 [see Table\,3 in Eracleous
\& Horne \cite{eh96}]. This means that the spin-down power
dominates the energy budget of the system and raises the question
about the nature of the spin-down torque, which is much larger
than any inferred accretion torque.

Some effort has been made to explain this phenomenon. Wynn, King
\& Horne \cite{wkh97} suggested that the white dwarf interacts
with the stream of material inflowing from the normal companion
and is spinning down due to propeller action. Ikhsanov \cite{i98}
has pointed out that the magneto-dipole losses of the white dwarf
are comparable with the estimated spin-down power provided its
surface magnetic field is $B(R_{\rm wd}) = 50 B_{50}$\,MG:
   \be\label{md}
L_{\rm md} = 5 \times 10^{33}\ B_{50}^2\ R_{8.8}^{6}\ P_{33}^{-4}\
{\rm erg\,s^{-1}},
   \ee
Here $R_{8.8}$ is the radius of the white dwarf expressed in units
of $10^{8.8}$\,cm, and the angle between the the rotational and
the magnetic axes of the white dwarf is taken according to
Eracleous \e \cite{erac94} as $\theta \approx 76^{\degr}$.

Both these approaches have been recently critically discussed by
Choi \& Yi \cite{cy00}, who presented an alternative explanation
of the spin-down power. Namely, they have shown that the rapid
braking of the white dwarf in \a\ can be explained in terms of the
gravitational wave emission mechanism provided the spatially
limited blobs or mounds of the mass
  \be\label{deltam}
\delta m \sim 10^{-2}\ [\sin^2{\theta}(13 \sin^2{\theta} +
\cos^2{\theta})]^{-1/2}\ {\rm M_{\sun}},
  \ee
are present at the magnetic poles of the white dwarf. Putting
$\theta=76^{\degr}$ we get
  \be\label{delm}
\delta m(\theta=76^{\degr}) \simeq 3\times 10^{-3} M_{\sun}.
  \ee

Choi \& Yi \cite{cy00} have assigned the origin of these mounds to
the accretion of material onto the white dwarf surface during the
previous spin-up epoch. They assumed that (i)\,the material
accreted by the white dwarf during this epoch impacted the stellar
surface preferentially in a local space at the magnetic pole
regions and (ii)\,the amount of this material, which was confined
by the white dwarf magnetic field at the magnetic pole regions,
was large enough for the mounds of the mass $\delta m$ to form.
Taking into account that the mass of the material accreted during
the spin-up epoch is limited to [see Ikhsanov \cite{i99}]
 \be
M_{\rm a0} \ga 3 \times 10^{-2}\,{\rm M_{\sun}\,yr^{-1}}\ I_{50}\
P_{33}^{-1}\ M_{0.8}^{-1/2},
 \ee
the second assumption implies that about 10\% of the accreted
material was stored in the mounds at the magnetic poles of the
white dwarf. Here $M_{0.8}$ denotes the mass of the white dwarf
expressed in units of $0.8 M_{\sun}$.

In this letter we analyze the validity of these assumptions. We
show that the size of the accretion region during the spin-up
epoch was almost comparable with the radius of the white dwarf and
that the maximum mass of the mounds, which can be confined by the
magnetic field of the white dwarf, is smaller than that estimated
from Eq.~(\ref{delm}) about 11 orders of magnitude.

   \section{The size of the accretion region}

As recognized by Lamb, Pethick \& Pines \cite{lpp73}, the size of
the accretion spot at the surface of a magnetized compact star can
be evaluated as
  \be\label{app}
a_{\rm p} \approx R_* \varepsilon_{\rm p}.
  \ee
Here $R_*$ is the radius of the star and $\varepsilon_{\rm p}$ is
the opening angle of the accretion column, which in the case of
the dipole configuration of the stellar magnetic field is
  \be
\varepsilon_{\rm p} \simeq (R_*/R_{\rm m})^{1/2},
  \ee
where $R_{\rm m}$ is the radius of the magnetosphere at the
magnetic equator.

For a steady accretion onto the stellar surface to occur the
radius of its magnetosphere must be limited to
  \be
R_{\rm m} \la R_{\rm cor} = 1.5 \times 10^9\ M_{0.8}^{1/3}\
P_{33}^{2/3} {\rm cm},
  \ee
where $R_{\rm cor}$ is the corotational radius and  $M_{0.8}$ is
the mass of the white dwarf expressed in units $0.8 M_{\sun}$.
Otherwise, the star is in the centrifugal inhibition regime and no
accretion onto its surface occurs [see for discussion Ikhsanov
\cite{i01}].

Under these conditions the opening angle of the accretion column
of the white dwarf during the previous spin-up epoch can be
limited to
  \be
\varepsilon_{\rm p0} \ga 0.65\ R_{8.8}^{1/2}\ \left[\frac{R_{\rm
m}}{1.5 \times 10^9\,{\rm cm}}\right]^{-1/2}\ {\rm rad},
  \ee
and, correspondingly, the size of the accretion region can be
estimated as [see Eq.~(\ref{app})]
  \be\label{ap}
a_{\rm p0} \approx  4.2 \times 10^8\ R_{8.8}^{3/2}\
\left[\frac{R_{\rm m}}{1.5 \times 10^9\,{\rm cm}}\right]^{-1/2}\
{\rm cm}.
  \ee

The obtained value of $a_{\rm p0}$ is only a factor of 1.5 smaller
than the radius of the white dwarf. In this case, the mounds
cannot be considered as spatially local blobs and hence, the
validity of the equation~(3.4) in Choi \& Yi \cite{cy00} is rather
questionable.

   \section{Upper limit to the mass of the mounds}

The stellar magnetic field is able to prevent the accreting plasma
from spreading around the stellar surface if the energy density of
the field is larger than the gas pressure in the mounds, i.e.
   \be
\frac{B_*^2}{8 \pi} > \frac{1}{2} \rho_{\rm m} V_{\rm s}^2.
  \ee
Here $B_*$ is the magnetic field strength in the magnetic pole
regions at the surface of the white dwarf, $\rho_{\rm m}$ is the
plasma density in the mounds, and $V_{\rm s}=\sqrt{\gamma
kT/m_{\rm p}}$ is the sound speed. $\gamma$, $k$ and $m_{\rm p}$
denote the adiabatic index, the Bolzmann constant and the proton
mass, respectively.

This means that the plasma density in the mounds can be limited to
  \be\label{rhom}
\rho_{\rm m} < \rho_{\rm max} \simeq 10^{-6}\ \mu_{32}^2\
R_{8.8}^{-6}\ T_8^{-1}\ {\rm g\,cm^{-3}},
  \ee
where $\mu_{32}$ is the magnetic moment of the white dwarf
expressed in units of $10^{32}\,{\rm G\,cm^3}$ and $T_8=T_{\rm
m}/10^8$\,K is the temperature in the shock at the base of the
accretion column, which is normalized according to the results of
Eracleous, Halpern and Patterson \cite{ehp91}. Hence, as soon as
$\rho_{\rm m}$ increases above $\rho_{\rm max}$, the magnetic
field can no longer balance the pressure of plasma in the mounds
and thus, the material accumulated in the magnetic pole regions is
able to spread around the surface of the white dwarf.

The maximum mass of mounds, which can be confined by the magnetic
field at the magnetic pole regions, can be estimated using
Eqs.~(\ref{ap}) and (\ref{rhom}) and assuming the radial
distribution of plasma density in the mounds to be barometric.
Under this assumption the maximum mass of mounds is
  \be\label{dmmax}
\delta m_{\rm max} \simeq \pi a_{\rm p}^2\ \int_0^{\rm r_{\rm
cor}}{\rho_{\rm max} e^{-(r/h_{\rm m}}) dr} \sim \pi a_{\rm p0}^2
\rho_{\rm max} h_{\rm m}
   \ee
   \bdm
\simeq 1.5\times 10^{-14}\ \mu_{32}^2\ M_{0.8}^{-1}\ R_{8.8}^{-1}\
\left[\frac{R_{\rm m}}{1.5 \times 10^9\,{\rm cm}}\right]^{-1}\
M_{\sun}.
   \edm
Here $h_{\rm m}=V_{\rm s}^2/g$ is the height of the homogeneous
atmosphere in the mounds and $g=GM_{\rm wd}/R_{\rm wd}^2$ is the
gravitational acceleration at the surface of the white dwarf.

Comparing Eqs.~(\ref{delm}) and (\ref{dmmax}), one can see that
for the mounds of the mass $3\times 10^{-3} M_{\sun}$ to be
confined by the magnetic field, the magnetic moment of the white
dwarf should be in excess of $4 \times 10^{37}\,{\rm G cm^3}$,
which implies the strength of the surface field of the star
$4\times 10^{10}$\,G. Under these conditions, however, the
magneto-dipole losses by the white dwarf would exceed the
evaluated spin-down power six orders of magnitude [see
Eq.~(\ref{md})].

On the other hand, if the magnetic moment of the white dwarf is
$\sim 10^{32}\,{\rm G\,cm^3}$, as adopted by Choi \& Yi
\cite{cy00}, the mass of mounds required to explain the observed
spin-down of the white dwarf within the gravitational wave emitter
mechanism is larger than the maximum possible mass of mounds,
which can be confined by the white dwarf magnetic field, by
roughly 11 orders of magnitude.

Even assuming that the radial distribution of plasma density in
the mounds is $\rho(r) \sim r^{-3}$, i.e. the same as the radial
distribution of the dipole magnetic field of the star, one finds
the maximum possible mass of mounds as $2\times
10^{-13}\,M_{\sun}$. This is still about ten orders of magnitude
smaller than that required within the approach of Choi \& Yi
\cite{cy00}.

Finally, the size of the accretion region, $a_{\rm p0}$, is
estimated under the assumption that the accretion flow onto the
surface of the white dwarf is steady and homogeneous. In the case
of inhomogeneous accretion, the size of the accretion region
depends on the parameters of blobs and may differ from the value
estimated by Eq.\,(\ref{ap}). However, if the radius of blobs is
larger than $a_{\rm p0}$, the size of the accretion region proves
to be comparable with the radius of the white dwarf. In this case
the equation\,(3.6) of Choi \& Yi \cite{cy00} turns out to be not
applicable. On the other hand, if the size of the accretion region
is smaller than that estimated in our paper, the density as well
as the gas pressure in the mounds of the same mass are larger.
Therefore, for these mounds to be confined, the magnetic field of
the white dwarf should be even stronger than that estimated above.

Thus, we are forced to conclude that the mounds of the mass
estimated by Eq.~(\ref{delm}) cannot exist on the surface of the
white dwarf and, hence, the approach presented by Choi \& Yi
\cite{cy00} is not applicable for the interpretation of the
observed rapid braking of the white dwarf in AE~Aquarii.

\acknowledgments

We would like to thank the anonymous referee for useful comments.
Nazar Ikhsanov acknowledge the support of the Alexander von
Humboldt Foundation within the Long-term Cooperation Program.

\end{document}